\documentclass[12pt]{iopart}

\usepackage{graphicx}
\usepackage{dcolumn}
\usepackage{bm}
\begin{document}
\title{CC1$\pi^+$ to CCQE cross sections ratio at accelerator energies}
\author{M. Sajjad Athar, S. Chauhan and S. K. Singh}
\address{Department of Physics, Aligarh Muslim University, Aligarh-202 002, India}
\ead{sajathar@rediffmail.com}
\begin{abstract}
The MiniBooNE and the K2K collaborations have recently reported the ratio for the $\nu_\mu$ induced charge current 1$\pi^+$ production cross section to the charged current induced quasielastic lepton production cross section i.e. $R(E)=\frac{\sigma^{CC1\pi^+}}{\sigma^{CCQE}}$. In this paper we study, theoretically, in a Local Density Approximation the effect of nuclear medium in charged current quasielastic(CCQE) and charged current one pion production(CC1$\pi^+$) processes for the $\nu_\mu$ induced reaction in nuclei. The theoretical results have been compared with the results of experimental observations by the MiniBooNE and the K2K collaborations. The ratio R(E) for $\nu_\mu$ induced reaction in water for the proposed T2K experiment is also presented. 
\end{abstract}
\section{Introduction}
Recently many dedicated accelerator neutrino experiments have been done both in the neutrino experiments energy region of around 1GeV~\cite{k2kcite} - \cite{minervacite} and at higher neutrino energies~\cite{minervacite} - \cite{novacite}. These accelerator experiments are looking for neutrinos in either a disappearance mode($\nu_l \not\rightarrow \nu_l$, l=$\mu$) or in an appearance mode ($\nu_l \rightarrow \nu_{l^\prime}$, l=$\mu$, $l^\prime=e$,$\tau$) to improve limit on the neutrino oscillation parameters of PMNS matrix~\cite{pmns1, pmns2}. It has been emphasized that a good knowledge of the cross section ratio is important in the analysis of $\nu_{\mu}$ disappearance searches where CC1$\pi^+$ events either contribute to signal events or to the large background to the CCQE signal. The MiniBooNE~\cite{mini1} and the K2K~\cite{k2k1} collaborations have reported the results for the observed $\frac{\sigma^{CC1\pi^+}}{\sigma^{CCQE}}$ cross section ratio R(E) as a function of neutrino energy. The K2K~\cite{k2k1} collaboration has also reported a flux integrated $\frac{CC1\pi^{+}}{CCQE}$ cross section ratio of $0.734 \pm 0.086 (fit)+^{+0.076}_{-0.103}(nucl.)+^{+0.079}_{-0.073}(sys.)$.  The MiniBooNE collaboration has presented their analysis in the neutrino energy region of 0.45GeV to around 2GeV for the detector using mineral oil(C$H_{2}$) while the K2K collaboration has presented their analysis for the SciBar(polystyrene $C_{8}H_{8}$) detector in the neutrino energy region of around 1 to 3 GeV. These measurements are meant for improving the limit on the CC1$\pi^+$ production cross section. The theoretical study of nuclear effect in the calculation of neutrino nucleus cross section becomes important in the $\nu_\mu$ induced charged current quasielastic lepton production(CCQE) as well as in the $\nu_\mu$ induced charged current 1$\pi^+$ (CC1$\pi^+$) production cross sections. The contribution to the CC1$\pi^+$ cross section comes from the incoherent as well as from the coherent neutrino induced pion production processes. In this paper, we have studied the nuclear effect in the neutrino nucleus cross sections in the energy region of around 1GeV and calculated the cross sections for the CCQE and CC1$\pi^+$ processes in the $\nu_\mu$ induced reaction in $^{12}C$ which is nuclear target in the MiniBooNE~\cite{mini1} and K2K~\cite{k2k1} experiments as well as in $^{16}O$ which is nuclear target in the proposed long baseline T2K~\cite{t2k} experiment.

The single pion production in this region is dominated by the resonance production in which a delta resonance is excited and decays subsequently to a pion and a nucleon. When this process takes place inside the nucleus, there are two possibilities, the target nucleus remains in the ground state leading to coherent production of pions or is excited and/or broken up leading to incoherent production of pions. We have considered both the production processes in the delta resonance model in the Local Density Approximation to calculate single pion production from nuclei. The effect of nuclear medium on the production of delta is treated by including the modification of delta properties in the medium. Once pions are produced, they undergo final state interactions with the final nuclei. For the incoherent pion production the final state interaction of the pions with the residual nucleus has been treated using a Monte Carlo code for pion nucleus interaction~\cite{Vicente}, while for the coherent case, the distortion of the pion plane wave is calculated by using the Eikonal approximation~\cite{Carrasco}. In the case of quasielastic reaction, the effect of Pauli principle and Fermi motion are included through the Lindhard function calculated in Local Density Approximation(LDA)~\cite{Singh1} and the modification in weak transition strengths in the nuclear medium has been taken into account through a Random Phase Approximation(RPA). The details of the formalism and calculations are already given in our earlier publications~\cite{Athar2} - \cite{Athar4}. Using the cross sections for the charged current quasielastic lepton production as well as the cross section for the charged current 1$\pi^+$ (CC1$\pi^+$) production processes in $^{12}$C and $^{16}O$, we obtain the ratio R(E) and compare our theoretical results with the experimental observations reported by the MiniBooNE~\cite{mini1} and the K2K~\cite{k2k1} collaborations and also compare the results from the earlier experiment performed at Argonne National Laboratory(ANL)~\cite{Radecky}, \cite{Barish}. The neutrinos in the MiniBooNE detector~\cite{MB_det} have been observed using a 12.2 m diameter spherical tank filled with 818 tons of mineral oil (CH$_2$) and the average energy of the muon neutrinos is around 750MeV. The K2K collaboration~\cite{k2k1} has reported the results for the neutrinos detected through a fully active scintillator detector, SciBar. SciBar consists of bars made of polystyrene (C$_8$H$_8$). The mean energy of the incident muon neutrinos is 1.3 GeV. The results of ANL experiment used here for the comparison were performed using Deuterium target in the Bubble chamber experiment~\cite{Radecky}. In this paper we also present the ratio $R(E)=\frac{\sigma^{CC1\pi^{+}}(E)}{\sigma^{CCQE}(E)}$ in water. These calculations have been performed for the proposed T2K experiment~\cite{t2k} which is a long baseline neutrino oscillation experiment using muon neutrinos produced by the decay of pions obtained from the protons at J-PARC. The average energy for the muon neutrinos is around 600MeV. A large water(H$_2$O) Cerenkov detector Super-Kamiokande is used as a far detector.
\section{QUASIELASTIC REACTION}
The basic reaction for the quasielastic process is a neutrino interacting with a neutron inside the nucleus i.e.
\begin{equation}\label{quasi_reaction}
\nu_{\mu}(k) + n(p) \rightarrow \mu^{-}(k^{\prime}) + p(p^{\prime})
\end{equation}
The cross section for quasielastic charged lepton production is calculated in the Local Density Approximation(LDA) as a function of local Fermi momentum of the nucleons inside the nucleus. In a nucleus, $\nu_\mu$ interacts with a neutron whose local density in the medium is $\rho_n(r)$ and is related to the local Fermi momentum as $p_{F_{n}}=[3\pi^2\rho_{n}(r)]^{\frac{1}{3}}$ where $\rho_{n}(r)$ is the neutron density taken from Ref.~\cite{Vries}. 

The Fermi motion and the Pauli blocking effect inside the nucleus have been taken into account through the imaginary part of the Lindhard function for the particle hole excitations in the nuclear medium~\cite{Singh1}. The total cross section $\sigma(E_\nu)$ for the charged current neutrino induced reaction on a nucleon inside the nucleus in a local Fermi gas model(FGM) is then written as~\cite{Athar2}:
\begin{eqnarray}\label{sigma_quasi}
\sigma_A(E)&=&-\frac{2{G_F}^2\cos^2{\theta_c}}{\pi}\int_{r_{min}}^{r_{max}} r^2 dr \int^{{k^\prime}_{max}}_{{k^\prime}_{min}} {k^\prime}^2dk^{\prime} \int_{-1}^1dcos\theta \frac{1}{E_{\nu_\mu} E_\mu} L_{\mu\nu} J^{\mu\nu}_{RPA} \nonumber\\
&&\times Im{U_N(q_0, {\bf q})}
\end{eqnarray}
where $L_{\mu\nu}=\sum L_\mu {L_\nu}^\dagger$ and ${J^{\mu\nu}_{RPA}}={\bar\Sigma}\Sigma J^\mu {J^\nu}^\dagger$, are calculated with RPA correlations in nuclei using leptonic current $L_\mu$ and the hadronic current $J^\mu$ given in Eq.(\ref{lep_curr}) below. $q(=k-k^\prime)$ is the four momentum transfer, M is nucleon's mass, $G_{F}$ is the Fermi coupling constant, $\theta_c$ is the Cabibbo angle and $U_N$ is the Lindhard function for the particle hole excitation~\cite{Singh1}.
The leptonic current $L_{\mu}$ and the hadronic current $J^\mu$ are given by
\begin{eqnarray}\label{lep_curr}
L_{\mu}&=&\bar{u}(k^\prime)\gamma_\mu(1-\gamma_5)u(k) \nonumber \\ 
J^\mu&=&\bar{u}(p^\prime)[F_{1}(q^2)\gamma^\mu + F_{2}(q^2)i{\sigma^{\mu\nu}}{\frac{q_\nu}{2M}} + F_{A}(q^2)\gamma^\mu\gamma_5 + F_{P}(q^2)q^\mu\gamma_5]u(p)
\end{eqnarray}
The form factors $F_i(Q^2) (i=1,2,A,P)$ are isovector electroweak form factors of the nucleon. We have taken the Bradford et al.~\cite{BBBA05} parametrization for the electromagnetic isovector form factors. The isovector axial form factor is taken to be of a dipole form with $M_A=1.1GeV$ and the pseudo-scalar form factor $F_p^V(Q^2)$ is given in terms of $F_A^V(Q^2)$ using the Goldberger-Treiman relation. The renormalization of weak transition strength in the nuclear medium in a random phase approximation(RPA) is taken into account by considering the propagation of particle hole(ph) as well as delta-hole($\Delta h$) excitations~\cite{Oset2}. These considerations lead to modified hadronic tensor components $(J^{\mu\nu}_{RPA})$ involving the bilinear terms in the electroweak form factors $F_i(Q^2) (i=1,2,A,P)$, for which expressions are given in Refs.~\cite{Athar2},~\cite{Nieves}.

\section{INCOHERENT PION PRODUCTION}
The basic reaction for the $\nu_\mu$ induced charged current inelastic one pion production process in nuclei is that a neutrino interacts with a nucleon N (proton or neutron) inside a nuclear target and the process is given by
\begin{equation}\label{chan_numu_pi+}\left.
\begin{array}{l}
\nu_\mu(k) + p(p)~\rightarrow~ \mu^{-}(k^\prime) + \Delta^{++}(P)\\
           ~~~~~~~~~~~~~~~~~~~~~~~~~~~~~~~~              \searrow p + \pi^+\\

\\

\nu_\mu(k) + n(p)~\rightarrow~ \mu^{-}(k^\prime) + \Delta^{+}(P)\\
           ~~~~~~~~~~~~~~~~~~~~~~~~~~~~~~~~                     \searrow p + \pi^0\\
           ~~~~~~~~~~~~~~~~~~~~~~~~~~~~~~~~                       \searrow n  + \pi^+\\
\end{array}\right\}
\end{equation} 
Here in our present work, we have taken $\Delta(1232)$ states and no higher $\Delta$ states have been taken into account. In the case of incoherent pion production in $\Delta$ dominance model, the weak hadronic currents interacting with the nucleons in the nuclear medium excite a $\Delta$ resonance which decays into pions and nucleons. The pions interact with the nucleons inside the nuclear medium before coming out. The combined final state interaction of pions through elastic, charge exchange scattering and the absorption of pions lead to reduction of pion yield. The nuclear medium effect on $\Delta$ properties lead to modification in its mass and width which have been discussed earlier by Oset et al.~\cite{Oset} and applied to explain the pion and electron induced pion production processes from nuclei. 
In the Local Density Approximation the expression for the total cross section for the charged current one pion production is written as
\begin{eqnarray}\label{sigma_inelastic}
\sigma_A(E)&=& \frac{1}{(4\pi)^5}\int_{r_{min}}^{r_{max}}(\rho_{p}(r)+\frac{1}{9}\rho_{n}(r)) d\vec r\int_{Q^{2}_{min}}^{Q^{2}_{max}}dQ^{2}\int^{{k^\prime}_{max}}_{{k^\prime}_{min}} dk^{\prime}\int_{-1}^{+1}dcos\theta_{\pi }\nonumber \\
&&
\times \int_{0}^{2\pi}d\phi_{\pi}\frac{\pi|\vec  k^{\prime}||\vec k_{\pi}|}{M E_{\nu}^2 E_{l}}\frac{1}{E_{p}^{\prime}+E_{\pi}\left(1-\frac{|\vec q|}{|\vec k_{\pi}|}cos\theta_{\pi }\right)}\bar\Sigma \Sigma|\mathcal M_{fi}|^2
\end{eqnarray}
The transition matrix element $\mathcal M_{fi}$ is given by
\begin{equation}\label{matrix_element}
\mathcal M_{fi}=\sqrt{3}\frac{G_F}{\sqrt{2}}\frac{f_{\pi N \Delta}}{m_{\pi}} \bar u({\bf p}^{\prime}) k^{\sigma}_{\pi} {\mathcal P}_{\sigma \lambda} \mathcal O^{\lambda \mu} L_{\mu} u({\bf p})
\end{equation}
In Eq.(\ref{sigma_inelastic}), a factor of $\frac{1}{9}$ comes with $\rho_n$ due to suppression of the production of $\pi^+$ from the neutron target ($\nu_\mu + n \rightarrow \mu^- + \Delta^+, \Delta^+ \rightarrow n + \pi^+) $ as compared to the $\pi^+$ production from the proton target through the process of delta excitation and decay in the nucleus. Here only $\Delta$(1232) excitations have been used. For our numerical calculations we take the proton density $\rho_{p}(r)=\frac{Z}{A}\rho(r)$ and the neutron density $\rho_{n}(r)=\frac{A-Z}{A}\rho(r)$, where $\rho(r)$ is nuclear density which we have taken as 3-parameter Fermi density given by 
\begin{equation}
\rho(r)=\rho_{0} \left(1 + w \frac{r^2}{c^2}\right) / \left(1 + exp\left(\frac{r - c}{z}\right)\right)
\end{equation}
and the density parameters c = 2.355 fm, z = 0.5224 fm and w = -0.149 for $^{12}C$ and c = 2.608 fm, z = 0.513 fm and w = -0.051 for $^{16}O$ are taken from Ref.~\cite{Vries}.

$L_{\mu}$ is the leptonic current defined by Eq.(\ref{lep_curr}) and $\mathcal O^{\lambda \mu}~(=~\mathcal O^{\lambda \mu}_V ~ + ~\mathcal O^{\lambda \mu}_V$) is the $N-\Delta$ transition operator. $\mathcal O^{\lambda \mu}_V$ is the vector part and $\mathcal O^{\lambda \mu}_A$ is the axial vector part of the transition operator and are given by
\begin{eqnarray}\label{mat_vector}
\mathcal O^{\lambda \mu}_V&=&\left(\frac{C^V_{3}(q^2)}{M}(g^{\alpha\mu}{\not q}-q^\alpha{\gamma^\mu}) + \frac{C^V_{4}(q^2)}{M^2}(g^{\alpha\mu}q\cdot{p^\prime}-q^\alpha{p^{\prime\mu}})+\frac{C^V_5(q^2)}{M^2}\right.\nonumber\\
&&\left.\times(g^{\alpha\mu}q\cdot p-q^\alpha{p^\mu})+ \frac{C^V_6(q^2)}{M^2}q^\alpha q^\mu\right)\gamma_5
\end{eqnarray}
and
\begin{eqnarray}\label{mat_axial.}
\mathcal O^{\lambda \mu}_A&=&\left(\frac{C^A_{3}(q^2)}{M}(g^{\alpha\mu}{\not q}-q^\alpha{\gamma^\mu}) + \frac{C^A_{4}(q^2)}{M^2}(g^{\alpha\mu}q\cdot{p^\prime}-q^\alpha{p^{\prime\mu}}) + C^A_{5}(q^2)g^{\alpha\mu}\right.\nonumber\\
&+&\left.\frac{C^A_6(q^2)}{M^2}q^\alpha q^\mu\right)
\end{eqnarray}
$C^V_i$(i=3-6) are the vector and $C^A_i$(i=3-6) are the axial vector N-$\Delta$ transition form factors. We have taken the parametrization of Lalakulich et al.~\cite{Lalakulich1} for the N-$\Delta$ transition form factors i.e.
\begin{equation}\label{civ_lala}
C_i^V(Q^2)=C_i^V(0)~\left(1+\frac{Q^2}{M_V^2}\right)^{-2}~{\cal{D}}_i~~,~~~i=3,4,5.
\end{equation}
where
\begin{eqnarray}\label{di} 
{\cal{D}}_i&=&\left(1+\frac{Q^2}{4M_V^2}\right)^{-1}~~~\mbox{for}~~~i=3,4~~~ and\nonumber\\
{\cal{D}}_i&=&\left(1+\frac{Q^2}{0.776M_V^2}\right)^{-1}~~~~~~\mbox{for}~~~i=5.
\end{eqnarray} 
and $C_3^V(0)=2.13, C_4^V(0)=-1.51$, $C_5^V(0)=0.48$ and the vector mass parameter $M_V$ = 0.84GeV. The conservation of charged vector current(CVC) implies $C_6^V(0)=0$.
The axial vector form factors are parametrized as~\cite{Lalakulich1}
\begin{eqnarray}\label{cia_lala}
C_i^A(Q^2)&=&C_i^A(0)~~\left(1+\frac{Q^2}{M_A^2}\right)^{-2}\left(1+\frac{Q^2}{3M_A^2}\right)^{-1},~~~i=3,4,5
\end{eqnarray}
where $C_3^A=C_4^A=0$, $C_5^A(0)=1.2$ and $C_6^A(Q^2) = M^2 \frac{C_5^A(Q^2)}{m_{\pi}^2 + Q^2}$, $Q^2=-q^2$, $M_A$ is the axial vector dipole mass and $m_\pi$ is the pion mass. For the numerical calculations we have taken $M_A$=1.1GeV.

In Eq.(\ref{matrix_element}), ${\mathcal P}^{\sigma \lambda}$ is the $\Delta$ propagator in momentum space and is given as~: 
\begin{equation}\label{width}
{\mathcal P}^{\sigma \lambda}=\frac{{\it P}^{\sigma \lambda}}{P^2-M_\Delta^2+iM_\Delta\Gamma}
\end{equation}
where ${\it P}^{\sigma \lambda}$ is the spin-3/2 projection operator given by
\begin{eqnarray}\label{propagator}
{\it P}^{\sigma \lambda} = \sum_{spins} \psi^{\sigma} \bar \psi^{\lambda} = (\not P+M_{\Delta})
\left(g^{\sigma \lambda}-\frac{2}{3} \frac{P^{\sigma}P^{\lambda}}{M_{\Delta}^2}+\frac{1}{3}\frac{P^{\sigma} \gamma^{\lambda}-P^{\sigma} \gamma^{\lambda}}{M_{\Delta}}-\frac{1}{3}\gamma^{\sigma}\gamma^{\lambda}\right)~~~~~~
\end{eqnarray}
and the delta decay width $\Gamma$ is taken to be a P-wave width~\cite{Oset}:
\begin{equation}\label{Width}
\Gamma(W)=\frac{1}{6 \pi}\left(\frac{f_{\pi N \Delta}}{m_{\pi}}\right)^2 \frac{M}{W}|{\bf q}_{cm}|^3
\end{equation}
$|{\bf q}_{cm}|$ is the pion momentum in the rest frame of the resonance and W is the center of mass energy.

Inside the nuclear medium, the mass and width of delta are modified as
\begin{equation}\label{modify}
\Gamma\rightarrow\tilde\Gamma - 2 Im\Sigma_\Delta~~and~~
M_\Delta\rightarrow\tilde{M}_\Delta= M_\Delta + Re\Sigma_\Delta.
\end{equation}
where $\tilde\Gamma$ is the Pauli reduced decay width. The expressions for $\tilde\Gamma$ and the real and imaginary part of the $\Delta$ self energy are taken from Oset et al.~\cite{Oset}. 

The pions produced in this process undergo multiple scattering and some of them are absorbed. This is treated in a Monte Carlo simulation which has been taken from Ref.~\cite{Vicente}. In the case of $\pi^+$, there is an additional channel available in a nuclear medium where its production may be enhanced. This is when a $\pi^0$ produced through the reaction $\nu_\mu + n \rightarrow \mu^- + \Delta^+, \Delta^+ \rightarrow p + \pi^0$ inside a nucleus may re-scatter through the nucleons present in the nuclei and give rise to a  $\pi^+$ through the strong interaction process i.e. $\pi^0 + p \rightarrow \pi^+ + n$. This has been taken into account by using the charge exchange scattering matrix discussed by Oset et al.~\cite{Oset3} and leads to a small increase of about 3-4$\%$ in the total $\pi^+$ production cross sections.
\section{COHERENT PION PRODUCTION}
The $\nu_{\mu}$ induced coherent one pion production on $^{12}C$ target i.e. $\nu_{\mu} + _{6}^{12}C \rightarrow \mu^{-} + _{6}^{12}C + \pi^{+}$ for which the cross section is given by an expression similar to Eq.(\ref{sigma_inelastic}). 
However, the matrix element $\mathcal M_{fi}$ is now given by 
\begin{equation}\label{matrix_coh}
\mathcal M_{fi} =\frac{G_{F}}{\sqrt{2}} cos\theta_{c} L^{\mu} J_{\mu} {\cal F}(\vec q - \vec k_{\pi})
\end{equation}
where $J_{\mu}$ the hadronic current is now given by 
\begin{equation}
J_{\mu}= \sqrt{3} \frac{f_{\pi N \Delta}}{m_{\pi}} \sum _{r,s} {\bar u_{s}}(p) k_{\pi \sigma} \mathcal P^{\sigma \lambda} \mathcal O_{\lambda \mu} u_{r}(p)
\end{equation}
${\cal F}(\vec q - \vec k_{\pi})$ is the nuclear form factor given by 
\begin{eqnarray}\label{ff}
{\cal F}(\vec q-\vec k_\pi)=\int d^{3}{\vec r} \left[{\rho_p ({\vec r})}+\frac{1}{3}{\rho_n ({\vec r})}\right]e^{-i({\vec q}-{\vec k}_\pi).{\vec r}}
\end{eqnarray}
When pion absorption effect is taken into account using the Eikonal approximation then the nuclear form factor ${\cal F}({\vec q}-{\vec k_\pi})$ is modified to $\tilde{\cal F}({\vec q}-{\vec k_\pi})$ which is calculated in Eikonal approximation to be~\cite{Athar4}:
\begin{eqnarray}\label{mod_ff}
\tilde{\cal F}({\vec q}-{\vec k_\pi})=2\pi\int_0^\infty b~db\int_{-\infty}^\infty dz~\rho({\vec b}, z)~J_0(k_\pi^tb) e^{i(|{\vec q}|-k_\pi^l)z} e^{-if({\vec b}, z)}
\end{eqnarray} 
where $
f({\vec b}, z)=\int_z^{\infty} \frac{1}{2|{\vec{k}_\pi}|}{\Pi(\rho({\vec b}, z^\prime))}dz^\prime$,
$k_{\pi}^{l}$ and $k_{\pi}^{t}$ are the longitudinal and transverse components of the pion momentum
and $\Pi$ is the self-energy of pion, the expression for which is taken from Ref.~~\cite{Oset}.
\section{RESULTS AND DISCUSSIONS}
Here we present and discuss the effects of nuclear medium on the total cross sections $\sigma$ for the $\nu_\mu$ induced charged current quasielastic lepton production process(CCQE), charged current incoherent one pion production process(Incoh-CC1$\pi^+$) and charged current coherent one pion production process(Coh-CC1$\pi^+$). Using them we obtain the ratio for the charged current one pion production(Incoherent+Coherent contributions) cross section to the charged current quasielastic lepton production cross section i.e. $R(E)=\frac{\sigma^{CC1\pi^{+}}(E)}{\sigma^{CCQE}(E)}$ and compared our numerical results with the recent measurements performed by the MiniBooNE~\cite{mini1} and the K2K~\cite{k2k1} collaborations and we also present the results for the planned T2K experiment~\cite{t2k}.
\begin{table*}
\begin{center}
\caption{Results for total cross section $\sigma$($\times10^{-38}cm^2$) for the $\nu_\mu$ induced charged current quasielastic lepton production process in $^{12}C$.}
\begin{tabular}{|c|c|c|c|c|c|}\hline\hline
$E_{\nu}$(GeV)& Free & Without RPA & With RPA\\ \hline\hline
0.15&0.65&0.08&0.04\\\hline
0.25&2.16&1.23&0.68\\\hline
0.45&4.62&3.73&2.69\\\hline
0.65&5.92&5.13&4.12\\\hline
0.85&6.51&5.78&4.87\\\hline
1.05&6.74&6.06&5.22\\\hline
1.25&6.82&6.17&5.36\\\hline\hline
\end{tabular}
\end{center}
\end{table*}

In Table-1, we present the results of $\sigma$ for the $\nu_\mu$ induced CCQE process obtained in $^{12}C$. These results have been presented for $\nu_\mu$ interacting with a free nucleon target as well as the CCQE process taking place from nucleons inside a Carbon target. The numerical results in $^{12}C$ have been obtained in the Fermi gas model using the Local Density Approximation with and without taking into account the nucleon correlation(RPA) effect. We find that when the cross section is calculated in the local Fermi gas model the reduction in the cross section is around 14$\%$ at $E_{\nu_\mu}$=0.65GeV and around $10\%$ at $E_{\nu_\mu}$=1GeV as compared to the the cross sections calculated for the free nucleon case. To check the consistency of our local Fermi gas model with the other Fermi gas models available in the literature, like the models of Llewellyn Smith~\cite{lsmith}, Smith \& Moniz~\cite{SM}, and Gaisser \& O'Connell~\cite{Gaisser}, we have found that our numerical results for the CCQE lepton production cross section are consistent with the results obtained by using other versions of the Fermi gas model~\cite{EPJA}. When we incorporate the RPA effect, there is a further reduction in the cross section which is more at lower neutrino energies and becomes smaller at higher neutrino energies, for example, the cross section further reduces by about $20\%$ at $E_{\nu_\mu}$=0.65GeV and around $14\%$ at $E_{\nu_\mu}$=1GeV. Thus, we find that the nucleon correlation effect plays an important role at the present neutrino energies of interest. 

In Table-2, we present the results for total cross section $\sigma$ for the $\nu_\mu$ induced incoherent charged current one pion production process(Incoh-CC1$\pi^+$) in $^{12}C$. The numerical results are presented for $^{12}C$  and have been obtained in the Local Density Approximation. First we obtain the cross section without taking into account the medium modification on the $\Delta$ properties which we call it as the calculations without medium effect(WME), then we take the modifications of the $\Delta$ properties in the nuclear medium using Eq.(\ref{modify}) and call it as the calculations with medium effect(ME). Our final results are those when we also take into account the final state interaction(FSI) of the outgoing pions with the residual nucleus(ME+FSI) following the method given in Ref.\cite{Vicente}. We find that the nuclear medium effect leads to a reduction of around 35$\%$ for $E_{\nu_\mu}$=1~GeV as compare to the cross section calculated without taking into account nuclear medium effect. When pion absorption effect is also taken into account there is a further reduction in the cross section which is around 16$\%$ for $E_{\nu_\mu}$=1~GeV. 
\begin{table*}
\begin{center}
\caption{Results for total cross section $\sigma$($\times10^{-38}cm^2$) for the $\nu_\mu$ induced incoherent charged current one pion production process in $^{12}C$.}
\begin{tabular}{|c|c|c|c|c|c|}\hline\hline
$E_{\nu}$(GeV)& Without Medium & With Medium & With ME + FSI \\
& Effect(ME) &  Effect(ME) &  Effect\\ \hline\hline
0.45&0.13&0.07&0.067\\\hline
0.65&1.58&0.89&0.76\\\hline
0.85&3.03&1.87&1.56\\\hline
1.05&4.05&2.62&2.21\\\hline
1.25&4.68&3.11&2.63\\\hline\hline
\end{tabular}
\end{center}
\end{table*}

In Table-3, we have presented the results for $\sigma$ for the $\nu_\mu$ induced coherent charged current one pion production process(Coh-CC1$\pi^+$) in $^{12}C$. The numerical results in $^{12}C$ have been obtained in the Local Density Approximation by taking the nuclear modification on the $\Delta$ properties and the final state interaction of pions with the residual nucleus. In this process the nuclear medium effect leads to a reduction of around 38$\%$ for $E_{\nu_\mu}$=1~GeV  as compared to the cross section calculated without taking into account the medium medium modification of $\Delta$-properties. When pion absorption effect is also taken into account there is a further reduction in the cross section which is around 50$\%$ for $E_{\nu_\mu}$=1GeV. 

\begin{table*}
\begin{center}
\caption{Results for total cross section $\sigma$($\times10^{-38}cm^2$) for the $\nu_\mu$ induced coherent charged current one pion production process in $^{12}C$.}
\begin{tabular}{|c|c|c|c|c|c|}\hline\hline
$E_{\nu}$(GeV)& Without Medium & With Medium & With ME + FSI \\
& Effect(ME) &  Effect(ME) &  Effect\\ \hline\hline
0.45&0.01&0.006&0.003\\\hline
0.65&0.126&0.07&0.034\\\hline
0.85&0.24&0.14&0.062\\\hline
1.05&0.36&0.22&0.11\\\hline
1.25&0.42&0.24&0.13\\\hline\hline
\end{tabular}
\end{center}
\end{table*}
Using the results of the total cross section $\sigma$ given in Tables 1-3, we obtain the ratio $R(E)=\frac{\sigma^{CC1\pi^{+}}(E)}{\sigma^{CCQE}(E)}$ and compare our numerical results with the recent measurements performed by the MiniBooNE~\cite{mini1} and the K2K~\cite{k2k1} collaborations. First we present and discuss the results obtained for the MiniBooNE experiment. For the Monte Carlo analysis of the neutrino events, the MiniBooNE collaboration has used v3 NUANCE event generator~\cite{NUANCE}. In the Monte Carlo simulation, CCQE interactions on carbon were obtained by using the relativistic Fermi gas model~\cite{SM}, \cite{MB_ccqe} with the axial dipole mass $M_A$=1.2GeV and the resonant CC$1\pi^+$ events were simulated using the Rein \& Sehgal (R-S) model \cite{RS} with an axial mass $M_A^{1\pi}=1.1$ GeV. Coherently produced CC$1\pi^+$ events were generated using the Rein \& Sehgal\cite{RS_coh} model with the R-S absorptive factor replaced by NUANCE's pion absorption model and the overall cross section rescaled to reproduce the MiniBooNE's measurement of neutral current coherent $\pi^0$ production \cite{MB_pi0}. 

At the MiniBooNE, in the mineral oil($CH_2$) there are two free protons, therefore, there would be charged current incoherent one pion production (Incoh-CC1$\pi^+$) from the free proton target as well, and total weight factor of $\frac{1}{3}$ would come with the free protons due to 6 protons and 6 neutrons inside the carbon nucleus.

\begin{table*}
\caption{Results for the inelastic charged current 1$\pi^+$ production cross sections without nuclear medium effect for the incoherent(Column-II) and coherent(Column-III) processes in $^{12}C$, inelastic production of pions from free proton target(Column-IV), charged current quasielastic lepton production cross section in the local Fermi gas model without RPA effect in $^{12}C$(Column-V) and the ratio $R(E)=\frac{\sigma^{CC1\pi^{+}}(E)}{\sigma^{CCQE}(E)}$ has been shown in Column-VI. }
\resizebox{148mm}{40mm}{
\begin{tabular}{|c|c|c|c|c|c|}\hline\hline
$E_{\nu}$(GeV)&Incoherent CC1$\pi^+$(without medium &Coherent CC1$\pi^+$(without medium &1/3 $\sigma$ (On free &Charged current quasielastic &$R(E)=\frac{\sigma^{CC1\pi^{+}}(E)}{\sigma^{CCQE}(E)}$)\\
& effect)& effect)&proton target)&process(Without RPA)&\\\hline\hline
0.45&0.13&0.01&0.0112&3.73&0.04\\\hline
0.55&0.77&0.06&0.053&4.55&0.19\\\hline
0.65&1.58&0.126&0.096&5.13&0.35\\\hline
0.75&2.35&0.188&0.137&5.52&0.48\\\hline
0.85&3.03&0.24&0.171&5.78&0.59\\\hline
0.95&3.59&0.28&0.2&5.95&0.68\\\hline
1.05&4.05&0.36&0.224&6.06&0.76\\\hline
1.15&4.4&0.40&0.243&6.13&0.82\\\hline
1.25&4.68&0.42&0.25&6.17&0.86\\\hline
1.35&4.88&0.44&0.267&6.18&0.90\\\hline\hline
\end{tabular}}
\end{table*}
In Table-4, using the results given in Tables-1-3, we present our results for the inelastic charged current 1$\pi^+$ production cross sections without nuclear medium effect for the incoherent(Column-II) and coherent(Column-III) processes in $^{12}C$, inelastic production of pions from free proton target(Column-IV), charged current quasielastic lepton production cross section in the local Fermi gas model without RPA effect in $^{12}C$(Column-V) and the ratio $R(E)=\frac{\sigma^{CC1\pi^{+}}(E)}{\sigma^{CCQE}(E)}$ is given in Column-VI. In Table-5, we present our results for the inelastic charged current 1$\pi^+$ production cross sections with nuclear medium and final state interaction effect for the incoherent(Column-II) and coherent(Column-III) processes in $^{12}C$, inelastic production of pions from free proton target(Column-IV), charged current quasielastic lepton production cross section in the local Fermi gas model with RPA effect in $^{12}C$(Column-V) and the ratio $R(E)=\frac{\sigma^{CC1\pi^{+}}(E)}{\sigma^{CCQE}(E)}$ is given in Column-VI. In Column-VI, we have also incorporated quasielastic like events, the contribution to which is coming from the inelastic channel when a pion does not appear in the final state and one only observes a lepton.
\begin{table*}
\caption{Results for the inelastic charged current 1$\pi^+$ production cross sections with nuclear medium and final state interaction effect for the incoherent(Column-II) and coherent(Column-III) processes in $^{12}C$, inelastic production of pions from free proton target(Column-IV), charged current quasielastic lepton production cross section in the local Fermi gas model with RPA effect along with quasilike events in $^{12}C$(Column-V) and the ratio $R(E)=\frac{\sigma^{CC1\pi^{+}}(E)}{\sigma^{CCQE}(E)}$ has been shown in Column-VI. }

\resizebox{148mm}{40mm}{
\begin{tabular}{|c|c|c|c|c|c|}\hline\hline
$E_{\nu}$(GeV)&Incoherent CC1$\pi^+$(with medium &Coherent CC1$\pi^+$(with medium &1/3 $\sigma$ (On free &Charged current quasielastic &$R(E)=\frac{\sigma^{CC1\pi^{+}}(E)}{\sigma^{CCQE}(E)}$)\\
&\& final state interaction effect)&\& final state interaction effect)&proton target)&process(With RPA+QE-like)&\\\hline\hline
0.45&0.067&0.003&0.0112&2.69 + 0.006 = 2.70&0.03\\\hline
0.55&0.356&0.015&0.053&3.50 + 0.055 = 3.55&0.12\\\hline
0.65&0.76&0.034&0.096&4.12 + 0.14 = 4.26&0.21\\\hline
0.75&1.17&0.05&0.137&4.56 + 0.23 = 4.8&0.28\\\hline
0.85&1.56&0.062&0.171&4.87 + 0.31 = 5.18&0.35\\\hline
0.95&1.91&0.08&0.2&5.09 + 0.37 = 5.46&0.40\\\hline
1.05&2.21&0.10&0.224&5.22 + 0.42 = 5.64&0.45\\\hline
1.15&2.45&0.11&0.243&5.31 + 0.45 = 5.76&0.49\\\hline
1.25&2.63&0.13&0.25&5.36 + 0.48 = 5.84&0.52\\\hline
1.35&2.76&0.14&0.267&5.39 + 0.50 = 5.89&0.54\\\hline\hline
\end{tabular}}
\end{table*}

\begin{figure*}
\begin{center}
\includegraphics[height=6cm, width=12cm]{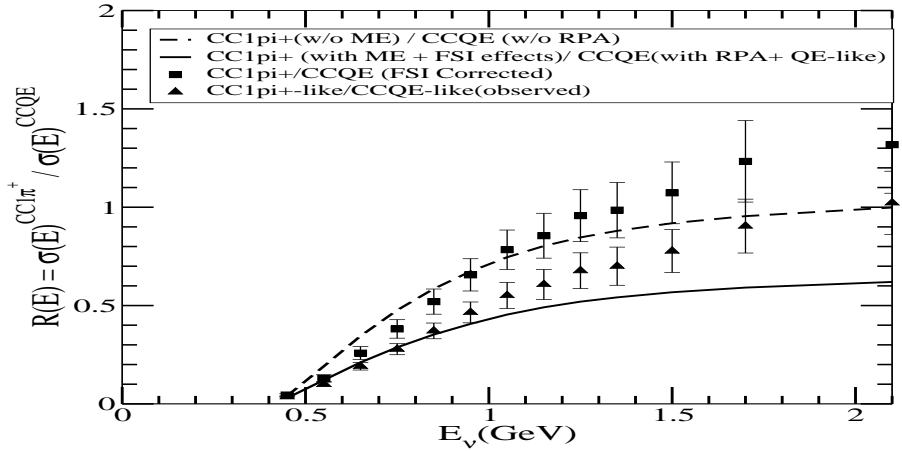}
\caption{Ratio of the cross sections for $\nu_{\mu}$ induced charged current one pion production process to the charged current quasielastic process in mineral oil. The dashed line shows the results corresponding to Column-VI of Table-1 and the solid line shows the results corresponding to Column-VI of Table-2. Squares shows the FSI corrected experimental points and triangle-up denotes the ratio of the cross sections for the observed events of $CC1\pi^+$-like/CCQE-like process in the MiniBooNE experiment~\cite{mini1}.}
\label{fg:Fig1_5}
\end{center}
\end{figure*}

Using the results of the cross sections shown in Tables-4 and 5, we have plotted in Fig.(\ref{fg:Fig1_5}) the ratio $R(E)=\frac{\sigma^{CC1\pi^{+}}(E)}{\sigma^{CCQE}(E)}$ of the cross sections for $\nu_{\mu}$ induced charged current one pion production process to the charged current quasielastic process. The one pion production includes contributions from incoherent as well as coherent channels. In the $\nu_{\mu}$ induced lepton production in $^{12}$C when the cross section for the pion production process is calculated without nuclear medium effect and the cross section for quasielastic lepton production process is calculated in the local Fermi gas model(FGM) without RPA effect, we find that the contribution from the inelastic channel is around 35$\%$ at $E_{\nu}$=0.65GeV and 75$\%$ at $E_{\nu}$=1GeV in comparison to the contribution of the lepton events from the charged current quasielastic process at these energies respectively.

Our final result for the ratio is the one where charged current one pion production cross section is calculated for $\nu_{\mu}$ induced reaction on free proton as well as from $^{12}$C nucleus with nuclear medium and final state interaction effect and the quasielastic lepton production cross section for $\nu_{\mu}$ induced reaction in $^{12}$C nucleus is calculated in the local Fermi gas model with RPA effect. This also includes the quasi-like events coming from the inelastic channel when a pion doesn't appear in the final state and one only observes a lepton. We find that the contribution from the inelastic channel is 20$\%$ at $E_{\nu}$=0.65GeV and around 42$\%$ at $E_{\nu}$=1GeV as compared to the contribution from the quasielastic channel.
In Fig.(\ref{fg:Fig1_5}), we compare our numerical results with the experimentally observed results and the FSI corrected results reported by the MiniBooNE collaboration~\cite{mini1}. FSI corrected results are those where using the Monte Carlo the final state interaction effect has been switched off i.e. no hadronic re-interactions have been considered. We find that our theoretical results for the ratio $R(E)=\frac{\sigma^{CC1\pi^{+}}(E)}{\sigma^{CCQE}(E)}$ obtained by switching off nuclear medium effect in the numerator and the denominator when calculated in the local Fermi gas model without RPA effect(Table-4) are in agreement with the FSI corrected results of the MiniBooNE~\cite{mini1}. When, in the ratio, $R(E)$ we consider the nuclear medium and final state interaction effect in the numerator and the denominator calculated in the local Fermi gas model with RPA effect along with the contribution of quasilike events from the inelastic channel(Table-5) the numerical results are in agreement with the experimental observed results reported by the MiniBooNE collaboration~\cite{mini1}. The agreement is better in the $\nu_\mu$ energy region of $E_{\nu_\mu}<$1.2 GeV.  Our results for R(E) are smaller than the experimental results for $E_{\nu_\mu}\geq$1.2 GeV. This may be due to the fact that at high neutrino energies the contribution to the 1$\pi^+$ cross section in nuclei is also expected to come from the higher nucleon resonances which has not been considered here. Our results of the cross sections calculated for CC1$\pi^+$ and CCQE reactions are in agreement with the recent calculations performed by various groups, like the calculations performed by Leitner et al.~\cite{Leitner}, Valverde et al.~\cite{Valverde} and Benhar et al.~\cite{Benhar} for the charged current quasielastic lepton production process, the calculations of Leitner et al.~\cite{Leitner} and Benhar et al.~\cite{Benhar} for the charged current incoherent pion production process and with the calculations of Alvarez-Ruso et al.~\cite{Coh-Ruso}, Leitner et al.~\cite{Coh-Leitner}, Amaro et al.~\cite{Coh-Amaro} and Berger and Sehgal~\cite{Coh-Berger} for the charged current coherent pion production process.

The K2K collaboration~\cite{k2k1} has reported the results for the single charged pion production in charged-current muon neutrino interactions with carbon using SciBar detector. The CC1$\pi$ production in the resonance region has been measured relative to the cross section of CCQE to reduce the impact of neutrino flux uncertainties. For the Monte Carlo analysis of the events K2K used the NEUT~\cite{Hayato} event generator. The K2K collaboration in their Monte Carlo analysis for the CCQE scattering used Smith \& Moniz~\cite{SM} Fermi gas model with Fermi surface momentum of 225~MeV for carbon and the axial dipole mass $M_{A}$=1.1 GeV.  For the production of single pions via baryon resonance excitations, Rein and Sehgal's model was used~\cite{RS, RS_1987} with axial mass $M_A^{1\pi}=1.1$ GeV. The effect of Pauli blocking in the resonance process and pion absorption effect has been taken into account~\cite{k2k1}.

\begin{figure}
\begin{center}
\includegraphics[height=6cm, width=12cm]{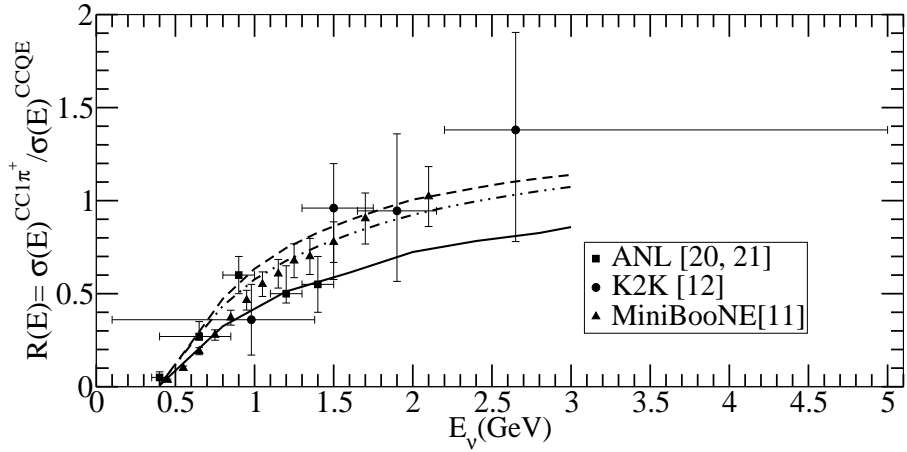}
\caption{$R(E)=\frac{\sigma^{CC1\pi^{+}}(E)}{\sigma^{CCQE}(E)}$ in polystyrene($C_{8}H_{8}$). The dashed(dashed-double dotted) line shows the results for R(E) obtained by using $\sigma^{CC1\pi^{+}}$  without(with) nuclear medium effect and $\sigma^{CCQE}$ in the FGM without RPA effect. Solid line shows our results for the cross section calculated with medium and FSI effect in the case of CC$1\pi^+$ production and quasielastic lepton production cross section calculated in FGM with RPA effect. The denominator also includes quasi-like events. Experimental points are the K2K(circle)~\cite{k2k1}, ANL(square)~\cite{Radecky}, \cite{Barish} and MiniBooNE(traingle-up) results~\cite{mini1}.}
\label{fg:Fig5}
\end{center}
\end{figure}

At K2K, in the  polystyrene (C$_8$H$_8$) target there are eight free protons and we calculate the charged current incoherent one pion production (Incoh-CC1$\pi^+$) from the free proton target as well and take a total weight factor of $\frac{1}{6}$ with free protons due to 48 protons and 48 neutrons inside the eight carbon nuclei. Using the numerical results given in Tables-1-3, we obtain the ratio $R(E)=\frac{\sigma^{CC1\pi^{+}}(E)}{\sigma^{CCQE}(E)}$ and plotted in Fig.(\ref{fg:Fig5}) our results for the ratio with and without nuclear medium effect. 

In Fig.(\ref{fg:Fig5}) using Tables-4 and 5, we show the results for the ratio of total one pion production to the quasielastic lepton production cross sections for $\nu_{\mu}$ induced reaction in polystyrene($C_{8}H_{8}$), i.e. $R(E)=\frac{\sigma^{CC1\pi^{+}}(E)}{\sigma^{CCQE}(E)}$ as a function of neutrino energy. Here we also compare our numerical results with the experimental observations of the  MiniBooNE~\cite{mini1} and the K2K~\cite{k2k1} collaborations and also with the earlier experimental results of ANL experiment which were performed using Deuterium target~\cite{Radecky}. To plot the different experimental results on the same figure MiniBooNE~\cite{mini1} and K2K~\cite{k2k1} results have been re-scaled to an isoscalar target by introducing a scaling factor of 0.80 for the MiniBooNE and 0.89 for the K2K~\cite{mini1} analysis. In the present analysis of ANL experiment no invariant mass cut has been applied while ANL used a cut on invariant mass $W <$ 1.4 GeV and the MiniBooNE spectrum is such that pion production events are considered only in the region $W <$ 1.6 GeV and for the K2K's measurement the invariant mass  $W <$ 2 GeV~\cite{k2k1}.

In this figure, we also show the ratio R(E)(dashed line) when no nuclear medium effect is taken into account either in the quasielastic process or in the inelastic process. When the cross sections are calculated without any nuclear medium modification except the Fermi motion and Pauli principle for the quasielastic process and without any modification of $\Delta$ properties in the nuclear medium in case of inelastic process, the results for R(E) are shown by dashed-double dotted line. We find that our numerical results for the ratio obtained by using local Fermi gas model(FGM) with RPA effect for the charged current quasielastic(CCQE) lepton production cross section and charged current one pion(CC1$\pi^{+}$) production cross section calculated with nuclear medium and final state interaction effect, are in agreement with the experimental observations for $E_{\nu_\mu}<$1.2 GeV.
 
Now we shall present the numerical results for the scattering cross section $\sigma$ obtained for $\nu_{\mu}$ induced charged current reaction in $^{16}O$ used in the calculation for R(E) obtained for water(H$_2$O) Cerenkov detector being used in the T2K experiment. The details of the numerical calculations may be found in Ref.\cite{EPJA}.

\begin{table*}
\caption{Results for the inelastic charged current 1$\pi^+$ production cross sections without medium effect(WME) and with nuclear medium(ME) effect for the incoherent(Column-II) and coherent(Column-III) processes in $^{16}O$, charged current 1$\pi^+$ production from free proton target(Column-IV), charged current quasielastic(CCQE) lepton production cross section(Column-V) in the local Fermi gas model without RPA effect in $^{16}O$ and the ratio $R(E)=\frac{\sigma^{CC1\pi^{+}}(E)}{\sigma^{CCQE}(E)}$ have been shown in Columns-VI \& VII. }

\resizebox{152mm}{10mm}{
\begin{tabular}{|c|c|c|c|c|c|c|}\hline\hline
$E_{\nu}$&{\large Incoherent CC1$\pi^+$}&{\large Coherent CC1$\pi^+$}&1/4 $\sigma$ (free&{\large CCQE }&$R(E)=\frac{\sigma^{CC1\pi^{+}}(E)}{\sigma^{CCQE}(E)}$)&$R(E)=\frac{\sigma^{CC1\pi^{+}}(E)}{\sigma^{CCQE}(E)}$\\
(GeV)&&&proton)&&&\\\hline\hline
\end{tabular}}

\resizebox{152mm}{30mm}{
\begin{tabular}{|c|c|c|c|c|c|c|c|c|c|}
& {\tiny WME} & {\tiny ME} & {\tiny WME} & {\tiny ME} &  & {\tiny Without}& &\\
& (I) & (II) & (III) & (IV) & (V) &  {\tiny RPA} (VI)& $\frac{(I+III+V)}{VI}$ &$\frac{(II+IV+V)}{VI}$\\\hline\hline
0.4&0.03&0.018&0.002&0.001&0.002&4.55&0.007&0.0046\\\hline
0.6&1.56&0.83&0.12&0.066&0.06&6.75&0.26&0.14\\\hline
0.8&3.59&2.12&0.29&0.17&0.12&7.82&0.51&0.31\\\hline
1.0&5.10&3.18&0.38&0.24&0.16&8.30&0.68&0.43\\\hline
1.2&6.12&3.95&0.49&0.31&0.20&8.48&0.80&0.52\\\hline
1.4&6.81&4.53&0.56&0.36&0.21&8.52&0.89&0.60\\\hline
1.6&7.32&4.97&0.58&0.40&0.23&8.50&0.96&0.66\\\hline\hline
\end{tabular}}
\end{table*}

In Table-6, we present our results for the inelastic charged current 1$\pi^+$ production cross sections without nuclear medium(WME) and with nuclear medium(ME) effect for the incoherent(Column-II) and coherent(Column-III) processes in $^{16}O$, inelastic production of pions from free proton target(Column-IV), charged current quasielastic lepton production cross section in the local Fermi gas model without RPA effect in $^{16}O$(Column-V). Using these results we obtain the ratio $R(E)=\frac{\sigma^{CC1\pi^{+}}(E)}{\sigma^{CCQE}(E)}$, where the numerator has been calculated without(Column VI) and with(Column VII) nuclear medium effect while the denominator has been calculated without RPA effect in our local Fermi gas model. The contribution from the inelastic channel for the $\nu_{\mu}$ induced lepton production in $^{16}$O when the cross section for the pion production process is calculated without nuclear medium effect and the cross section for quasielastic lepton production process is calculated in the local Fermi gas model without RPA effect is around 68$\%$ at $E_{\nu}$=1GeV in comparison to the contribution of the lepton events from charged current quasielastic process. With the incorporation of nuclear medium effect in the inelastic one pion production process this contribution becomes around 42$\%$ at $E_{\nu}$=1GeV when compared with the contribution of the quasielastic lepton production calculated in the local Fermi gas model without the RPA effect.

\begin{table*}
\caption{Results for the inelastic charged current 1$\pi^+$ production cross sections with nuclear medium and final state interaction effect for the incoherent(Column-II) and coherent(Column-III) processes in $^{16}O$, inelastic production of pions from free proton target(Column-IV), charged current quasielastic lepton production cross section in the local Fermi gas model with RPA effect in $^{16}O$(Column-V) and the ratio $R(E)=\frac{\sigma^{CC1\pi^{+}}(E)}{\sigma^{CCQE}(E)}$ has been shown in Column-VI. }

\resizebox{152mm}{40mm}{
\begin{tabular}{|c|c|c|c|c|c|}\hline\hline
$E_{\nu}$(GeV)&Incoherent CC1$\pi^+$(with medium &Coherent CC1$\pi^+$(with medium &1/4 $\sigma$ (On free &Charged current quasielastic &$R(E)=\frac{\sigma^{CC1\pi^{+}}(E)}{\sigma^{CCQE}(E)}$)\\
&\& final state interaction effect)&\& final state interaction effect)&proton target)&process(With RPA+QE-like)&\\\hline\hline
0.4&0.017&0.0008&0.002&3.70 + 0.001 = 3.70&0.005\\\hline
0.6&0.70&0.03&0.06&5.90 + 0.12 = 6.02&0.13\\\hline
0.8&1.76&0.08&0.12&7.06 + 0.35 = 7.41&0.26\\\hline
1.0&2.66&0.12&0.16&7.60 + 0.52 = 8.12&0.36\\\hline
1.2&3.34&0.16&0.20&7.80 + 0.62 = 8.42&0.44\\\hline
1.4&3.85&0.18&0.21&7.86 + 0.68 = 8.54&0.50\\\hline
1.6&4.25&0.21&0.23&7.85 + 0.72 = 8.57&0.55\\\hline\hline
\end{tabular}}
\end{table*}

\begin{figure*}
\begin{center}
\includegraphics[height=6cm, width=12cm]{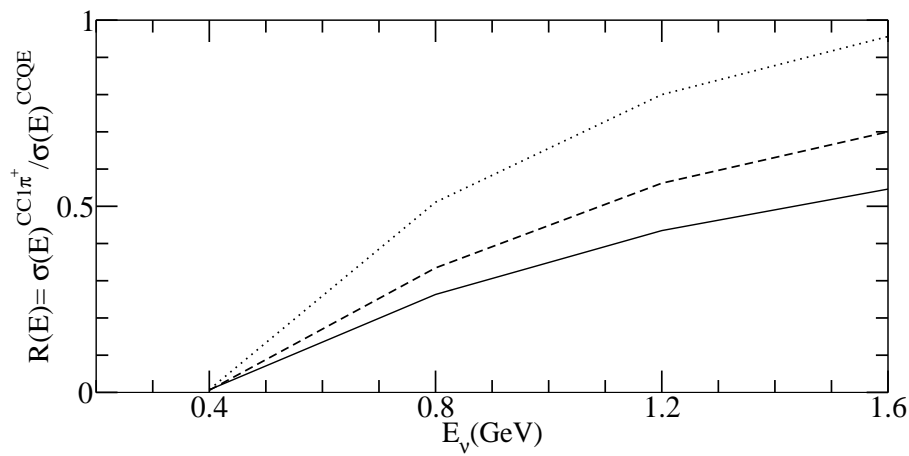}
\caption{Ratio $R(E)=\frac{\sigma^{CC1\pi^{+}}(E)}{\sigma^{CCQE}(E)}$ of the cross sections for $\nu_{\mu}$ induced CC1$\pi^+$ production process to the CCQE lepton production process. The dotted(dashed) line shows the ratio without(with) nuclear medium effect in the case of 1$\pi^+$ production process and the quasielastic lepton production cross section is calculated in FGM without(with) RPA effect. The solid line is the ratio when nuclear medium effect and the FSI effect is taken into account in the case of 1$\pi^+$ production process and the quasielastic lepton production cross section is calculated in FGM with RPA effect and it also includes quasielastic like events.}
\label{fg:Fig17}
\end{center}
\end{figure*}

In Table-7, we present our final results for the ratio R(E) when the inelastic charged current 1$\pi^+$ production cross section is obtained with nuclear medium and final state interaction effect for the incoherent(Column-II) and coherent(Column-III) processes in $^{16}O$, inelastic production of pions from free proton target(Column-IV) and the charged current quasielastic lepton production cross section is obtained in the local Fermi gas model with RPA effect(Column-V). The ratio $R(E)=\frac{\sigma^{CC1\pi^{+}}(E)}{\sigma^{CCQE}(E)}$ is given in Column-VI. In Column-VI, we also incorporate quasielastic like events, the contribution of which is coming from the inelastic channel when a pion does not appear in the final state and one only observes a lepton. We find that the contribution from the inelastic channel is around 14$\%$ at $E_{\nu}$=0.6GeV and 36$\%$ at $E_{\nu}$=1GeV as compared to the contribution from the quasielastic channel. Using the results given in Tables-6 and 7, we have plotted the ratio $R(E)=\frac{\sigma^{CC1\pi^{+}}(E)}{\sigma^{CCQE}(E)}$ in Fig.(\ref{fg:Fig17}) and find that the nuclear medium effect is important both for charged current quasielastic lepton production as well as charged current 1$\pi^+$ production processes for the T2K energies.

\begin{figure*}
\begin{center}
\includegraphics{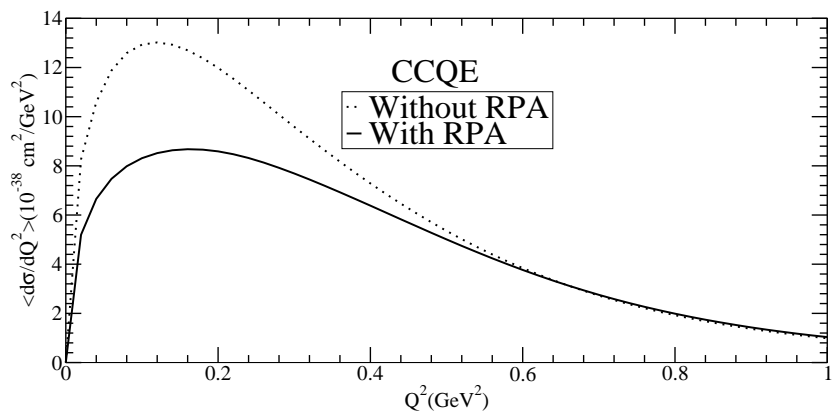}
\caption{$Q^2$ distribution for $\nu_{\mu}$ induced charged current quasielastic process in $^{16}O$ averaged over the T2K spectrum.}
\label{fg:Fig13}
\end{center}
\end{figure*}

\begin{figure*}
\begin{center}
\includegraphics{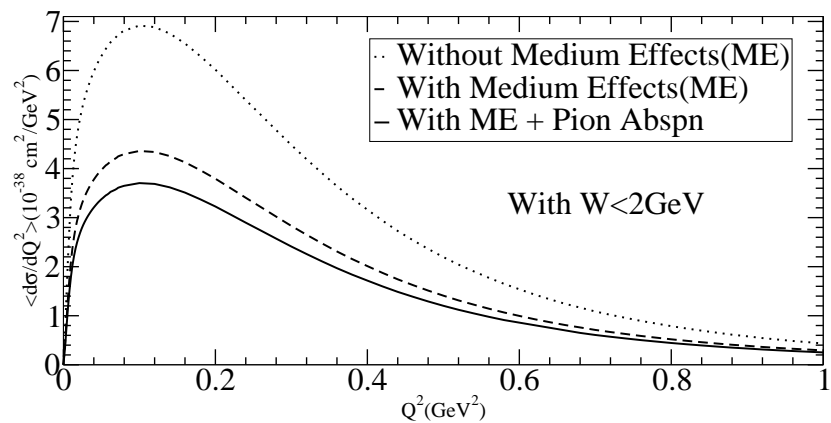}
\caption{$Q^2$ distribution for $\nu_{\mu}$ induced charged current incoherent 1$\pi^+$ production process in $^{16}O$ averaged over the T2K spectrum.}
\label{fg:Fig15}
\end{center}
\end{figure*}

To examine the dependence of $Q^2$ distribution on the nuclear effect, we study this distribution for the $\nu_\mu$ induced reaction in $^{16}O$ averaged over the T2K flux for CCQE and CC1$\pi^+$ processes. In Fig.(\ref{fg:Fig13}) the $Q^2$ distribution is shown for the charged current quasielastic process. We find that the reduction in the $Q^2$ distribution with RPA effect taken into account is around 35$\%$ in the peak region ($Q^2$ = 0.12$GeV^2$) which becomes 6$\%$ at $Q^2$ = 0.5$GeV^2$. In Fig.(\ref{fg:Fig15}) we have shown the $Q^2$ distribution for the incoherent charged current one pion production process induced by $\nu_\mu$ averaged over the T2K flux. We find that the reduction in the $Q^2$ distribution when medium effect is taken into account is around 35$\%$ which gets further reduced by 15$\%$ when we incorporate pion absorption effect.

\section{Concluding Remarks}
To conclude, we have studied in this paper the effect of nuclear medium on the inclusive quasielastic and charged current one pion production in nuclei and compared our results, for the ratio $R(E)=\frac{\sigma^{CC1\pi^{+}}(E)}{\sigma^{CCQE}(E)}$, with the recent experimental results from the MiniBooNE~\cite{mini1} and K2K~\cite{k2k1} experiments. We have also applied our results to calculate the ratio R(E) for $\nu_\mu$ induced reaction in water relevant for the T2K experiment which is planning to observe $\nu_\mu$ disappearance using water Cerenkov detector. We find that the nuclear medium effect is important in the energy region of present accelerator experiments discussed here and the ratio R(E) depends significantly on the nuclear effect in the CCQE as well as $CC1\pi^+$ processes.\\

{\bf Acknowledgments}\\

The work is supported by the DST, Government of India under the grant SR/S2/HEP-0001/2008. \\

{\bf References}\\

\end{document}